# Towards a Model of Systemic Change in University STEM Education

Daniel L. Reinholz*, Joel C. Corbo*, Melissa H. Dancy**, Noah Finkelstein**,
and Stanley Deetz†

*Center for STEM Learning, University of Colorado, Boulder, CO 80309
**Department of Physics, University of Colorado, Boulder, CO 80309
†Graduate School, University of Colorado, Boulder, CO 80309

*Despite numerous calls for the transformation of undergraduate STEM education, there is still a lack of successful models for creating large-scale, systemic cultural changes in STEM departments. To date, change efforts have generally focused on one of three areas: developing reflective teachers, disseminating curricula and pedagogy, or enacting institutional policy. These efforts illustrate many of the challenges of departmental change; in particular, they highlight the need for a holistic approach that integrates across all three of these levels: individual faculty, whole departments, and university policymakers. To address these challenges, we import and integrate models of change from multiple perspectives as part of our STEM education transformation effort. We draw from models in organizational change, from departmental and disciplinary change in STEM education, and from efforts to support individuals such as through the development and dissemination model. As a result, our departmental cultural change efforts are an attempt at holistic reform. We will discuss our theoretical underpinnings and ground this theory in a sample of approaches in two departments*

## Introduction

Converging evidence concludes that certain types of teaching practices are most likely to improve student outcomes in undergraduate STEM courses (e.g., Freeman et al., 2014). Despite efforts to document and disseminate such practices, they are still not widely adopted (Henderson & Dancy, 2009). This lack of adoption suggests the need for new models and approaches towards institutional change (PCAST, 2012). This paper advances such a model and describes our approaches to implementing it.

## Institutional Change in Higher Education

While institutional change models are well-developed in business and government settings (Real & Poole, 2005), similar models in higher education are only beginning to emerge (e.g., AAAS, 2011; Chasteen, Perkins, Beale, Pollock, & Wieman, 2011; Henderson, Beach, & Finkelstein, 2011). These emergent models focus primarily on changing *practices* associated with teaching and learning. Our model builds on these efforts by adding a focus on culture in addition to practice, which we argue is required to effect sustained change. We situate our model with respect to prior efforts, particularly the Science Education Initiative (SEI), and highlight points of divergence.

The SEI is a course transformation effort aimed at STEM departments across two institutions (Chasteen et al., under review, 2011), one of which we are presently working with. The SEI focuses on transforming individual courses across a department using a three-component model: (1) defining learning goals for a course, (2) identifying areas of student difficulty, and (3) developing materials to help students meet the now-established learning goals. Science Teaching Fellows, disciplinary experts with educational training, were hired into each department to help promote and guide this transformation process. It typically took two to three semesters to develop learning goals collaboratively and to implement and refine new instructional approaches. While the SEI is largely considered successful, it did not explicitly focus on changing culture within departments. With respect to this, our faculty interviews have provided evidence of "slippage" in departments where reforms were made, due to the end of funding and new faculty who were not involved with the SEI teaching the transformed courses. In our work with such departments, we take the positive impact of SEI as a starting point for our own change efforts, with careful attention to sustaining and improving reforms.

Our change model is built on the following principles, elaborated below:

1. We focus on both prescriptive and emergent components of change (Henderson et al., 2011).
2. We pay explicit attention to sustaining the change process, focusing on continued improvement (Phillips, 1977).
3. We recognize the existing culture and institutional constraints, while focusing on reforming incentive structures to seed and sustain change (AAAS, 2011).
4. We take the department as a key unit of change (AAAS, 2011), while recognizing the need to target the university ecosystem at multiple levels.

These principles derive from and reflect findings from prior STEM educational transformation efforts. Henderson et al. (2011) classify such efforts across two dimensions: the primary target of change (individuals vs. environments) and whether the outcome was known in advance (prescribed vs. emergent). The dissemination of best practice curricula (prescribed-individuals) and top-down policy making (prescribed-environments) were found to be ineffective in isolation. For instance, efforts targeted at faculty can be limited by a highly traditional environment (Henderson & Dancy, 2007). These findings suggest that change efforts should target the university at multiple levels (Henderson et al., 2011).

Our experiences with the SEI highlight the need to continuously sustain change efforts. In general, efforts heavily driven by external support tend to regress after the support is removed (i.e., education problems do not stay solved; Phillips, 1977), unless there are structural changes that are difficult to reverse (e.g., replacing lecture halls with SCALE-UP style, or studio, classrooms; cf. Beichner et al., 2007). In contrast, efforts that result in cultural change, rather than just shifts in practice, may help a department sustain its efforts without continued external support. Nevertheless, fundamental changes to institutional reward structures, providing adequate incentives for improving teaching and learning, are key to sustaining reform efforts (AAAS, 2011).

We take the department as a key unit of change; this allows for efforts to be integrated across the curriculum, rather than being implemented piecemeal in isolated courses (AAAS, 2011). Given the complexity of the university (Henderson et al., 2011), change strategies focused solely on individuals, not the systems they are embedded in, are unlikely to succeed; our model addresses the university ecosystem at multiple levels.

**Core Commitments**

Our change model targets the development of departmental cultures that are aligned with six core commitments, which we believe are emblematic of the culture of highly effective departments that value undergraduate education:

1. Educational experiences are designed around clear learning outcomes.
2. Educational decisions are evidence-based.
3. Active collaboration and positive communication exist within the department and with external stakeholders.
4. The department is a "learning organization" focused on continuous improvement.
5. Students are viewed as partners in the education process.
6. The department values inclusiveness, diversity, and difference.

Although academic departments are not typically viewed as networks, they share important characteristics with Networked Improvement Communities (NICs; Bryk, Gomez, & Grunow, 2011). NICs are collaborative networks organized to address complex, persistent problems in education. Departments also attempt to address such problems; this requires: (1) a clear goal, (2) gathering of evidence to evaluate proposed solutions, (3) mechanisms for positive coordination and collaboration, and (4) mechanisms for sustained learning and improvement (Bryk et al., 2011). These four statements, aligned with our first four core commitments, establish a functional problem-solving process. Moreover, addressing complex problems requires diverse skills and perspectives (Bryk et al., 2011), so the perspectives of students (commitment 5), particularly those from diverse backgrounds (commitment 6), are of significance.

**A Model for Institutional Change**

Our change model addresses the university system as a whole, with careful attention to connections across three levels: faculty (as individuals and groups), departments, and administration. Activities at each level are synergistic; changes in a department influence both individual faculty and the administration (our middle-out approach), and efforts at the outside levels (faculty and administration) influence the department itself (our outside-in approach). Our two approaches (see Figure 1) are synergistic, and most effective when used together. Nevertheless, we have begun by studying and implementing these approaches separately, to better develop theory and practice before using them together in an integrated effort.

Figure 1. The outside-in approach (blue arrows) involves providing external support to the administration and faculty to impact the department. The middle-out approach (orange arrows) focuses on the department directly. Because all levels of the university are linked, both approaches are intended to affect the university at all levels.

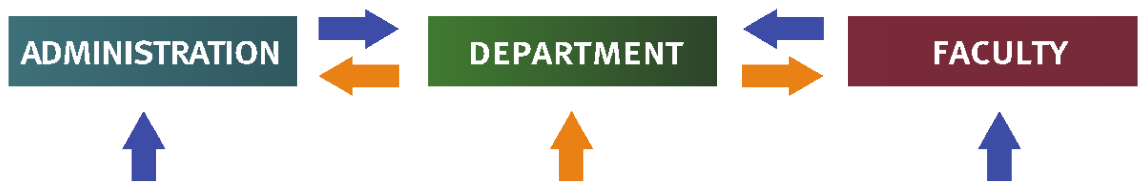

*The Outside-In Approach*

The outside-in approach combines efforts at the faculty level and administrative level to shift department culture. By working with groups of faculty, as in the SEI (Chasteen et al., 2011), we aim to reform educational practices and shift beliefs about education. Contrasting with prior approaches (Henderson et al., 2011), our efforts focus explicitly on cultural change.

Our approach involves creating Departmental Action Teams (DATs), which consist of faculty working collaboratively to address a shared issue of departmental interest. Like a faculty learning community (FLC; Ortquist-Ahrens & Torosyan, 2009), DAT faculty have agency to choose the educational issue they will work on, and *learning* and *community* are considered central to the DAT. DATs differ from FLCs insofar as they focus on a common, shared issue in a single department. Our DATs embody our six core commitments (e.g., use of evidence, clear outcomes) and engage faculty in a collaborative process aimed at shifting how faculty engage in scholarship of teaching and learning (SoTL; Huber & Hutchings, 2005); We see DATs as a mechanism for local cultural change. To sustain our efforts, DATs must be institutionalized through departmental support.

At the administrative level, the outside-in approach focuses on shifting university incentive structures and resources. At most research universities, there is little incentive for faculty to invest the time and effort required to teach effectively, because such investment is viewed as conflicting with research productivity. When individuals (or even collections of individuals) do engage, they often do so in isolation, developing their own tools (Glassick, Huber, & Maeroff, 1997). Thus, to sustain educational transformation, shifts in institutional incentive structures and resources are required. Our approach involves working with a variety of institutional structures to: (1) prioritize research-based teaching practice and (2) provide resources for faculty, departments, and administration to do so. Similarly, working with administration we seek to promote a faculty-developed framework for defining (and celebrating) teaching excellence that can be adopted and contextualized by individual departments. Such a framework and approach could provide needed, locally relevant tools to shift the promotion and tenure guidelines and culture, if promoted by institutional leadership (cf. Iowa State University, 2014). In parallel with efforts to promote effective educational practices (e.g., technological tools, easily accessed data on educational outcomes, and coordinated support for pedagogical development; cf. Berret, 2014), we seek to simultaneously provide incentives and resources that promote a culture of educational excellence.

*The Middle-Out Approach*

The middle-out approach involves a sustained change process focused on aligning department culture with our core commitments. We adopt strategies from the organizational change literature (e.g., Conversant, 2014) that have been successful in systems similar to academic departments to guide departments through a process involving five inter-related components:

1. developing a department vision,
2. revising assumptions about teaching and learning,
3. developing capacity to meet learning goals (within and across courses),
4. integrating teaching and learning goals systematically with research and other departmental functions, and
5. developing a collaborative process for continuous assessment and innovation.

We take our six core commitments above as a starting point to creating a shared vision, but allow for the department to build upon and interpret them within their local context. In essence, our core commitments lay out basic parameters for what a collaborative process might achieve. Through individual faculty interviews, we create "mental maps" of their beliefs around teaching and learning. The maps help identify areas for productive change efforts and also barriers to faculty embracing the shared vision (Borrego & Henderson, 2014). The mental maps are also tools for intervention: by sharing them with the faculty, we can shift faculty beliefs towards alignment with the vision by revealing assumptions and incongruities of which the faculty were not previously aware. Steps (3)-(5) of our process involve the department building capacity, integrating meaningful teaching and learning across all department activities, and developing collaborative processes for ongoing assessment and innovation.

The activities associated with these change processes: increase awareness of competing values, inculcate more productive maps of faculty work and student learning, and increase department capacity to meet multiple goals. As a result, appropriate student outcomes emerge without being prescribed by the change process as faculty shift their practices to align with the new culture they are co-creating. This shift in culture also has the potential to impact the administration (e.g., demonstrated success of the process could convince the administration to encourage other departments to engage in a similar process). These changes at the faculty and administration level can then further influence department culture, creating a sustained feedback cycle.

We anticipate this change effort would last 1-2 years. The process would involve, at minimum, a 1-2 day departmental retreat to develop a vision, mental maps, assessment criteria, and a process going forward that includes 30-day, 90-day, and one-year goals. The retreat would establish working groups to complete different tasks (e.g., establishing learning goals for the program, creating a more supportive environment for innovations and positive relationships, and revising reward systems). At regular intervals, the department would meet to assess progress, reflect on successes and lessons learned, and adjust its process to move forward. The members of our project team will be involved as facilitators of this change process.

**Sample Applications**

Our preliminary efforts to implement these two approaches offer several insights. We are currently engaged in our second year of activities associated with the STEM Institutional Transformation Action Research (SITAR) Project. SITAR is a three-year grant-funded project to implement and study institutional change at a large research university in the USA. Our project is presently involved with four STEM departments, with plans to scale up over time. For this brief paper, we report on our efforts with two departments: the Runes Department and the Charms Department (actual names redacted for confidentiality). These case studies highlight our approaches. Both of these departments were prior recipients of funding through the SEI.

In all departments involved in the project we administered a survey focused on teaching practices, beliefs, professional development opportunities, and promotion and tenure guidelines (all measured on Likert scales). In the Runes Department, we conducted nine, one-hour interviews with individual faculty about their teaching and their perceptions of the department; we also engaged in one-on-one consultations with two faculty around education projects. In the Charms Department we conducted a faculty survey and held several meetings with departmental leadership to determine readiness for engagement in our change processes.

*The Runes Department: The Outside-In Approach*

In the Runes Department, 31 of 34 faculty responded to our survey, for a response rate of 91%. Faculty indicated that social interaction (4.25 out of 5), active participation (4.75 out of 5), interactive learning (4.66 out of 5) and engagement (4.1 out of 5) were very important aspects of learning. Despite these beliefs, use of small group work (2.6 out of 4), regular opportunities for students to talk (2.8 out of 4), and opportunities for students to explore content before formal instruction (2.3 out of 4) were mixed. Most telling, instructors indicated that the statement "I guide students through major course topics as they listen and take notes" was mostly descriptive of their teaching (3.2 out of 4). These survey responses seem to indicate a discontinuity between professed beliefs and actual teaching practices, indicative of the institutional culture and incentive structures. This may be the result of SEI's focus on the practices of faculty in this department with no corresponding changes at the administrative level.

Runes is spread across multiple buildings on two campuses. As a result, many of the faculty we interviewed reported feeling isolated; through the creation of a DAT, we aimed to create more community around education. Our DAT consists of six faculty members (a mix of tenure-track and non-tenure track faculty), and two facilitators from our project team; many of the DAT members were identified through our individual faculty interviews. Our DAT began this fall and will continue meeting regularly through the spring. The DAT has received department support, in the form of an instructor course buyout for the department lead, service credit for all members, and the sanction of the department chair and teaching committee.

In the outside-in approach, our work at the administrative level is intentionally lagging the work at the departmental level. At the administrative level, the senate has shifted its calls in its campus-wide Teaching Awards to focus on evidence-based practices (University of Colorado, 2008). Additionally, working with the faculty senate we have revised the tools for evaluating "excellence" in teaching awards (e.g., to include evidence of scholarship in teaching and learning, measures of student learning and engagement; cf. University of Colorado, 2013). Once this framework has been established and interpreted by departments and once the DATs have begun the process of normalizing evidence-based conversations about education, we will then work with the senior administration to require evidence of student learning in tenure and promotion decisions. In this sense, "top-down" efforts are phase-shifted relative to our "bottom-up" work; only once there is sufficient faculty buy-in would such mandates be implemented. This would serve to institutionalize the changes already happening at the departmental level.

*The Charms Department: The Middle-Out Approach*

In the Charms Department, thirteen faculty members responded to a survey gauging interest in participating in our cultural change process. Twelve of these faculty indicated that they were interested or very interested, with only one faculty member reporting no interest. Based on this survey, the decision to participate in our process was brought to a faculty vote; faculty unanimously agreed to participate. We have begun our preliminary data collection efforts this fall, and will begin the change process in the spring.

To measure the impact of our process, we are currently revising the PULSE vision and change rubrics (Bianco, Jack, Marley, & Pape-Lindstrom, 2013) to better capture shifts in culture, not just practice. We also intend to use surveys at the individual faculty level (inspired by Henderson, under development). The PULSE will serve both as a measurement tool (for pre/post testing department culture) and as a formative intervention tool for facilitating discussions around cultural change.

**Conclusion**

There is an urgent need for cultural change in STEM departments. Our change model aims to effect cultural change in alignment with six core commitments for productive departmental culture. We provide two synergistic approaches to using the model, which address the university ecosystem as a whole. Ultimately, these two approaches should be used in conjunction to effect systemic, sustained reform in STEM departments. These approaches may be used simultaneously or sequentially; for instance, initial efforts and success with the outside-in approach might prompt a department to seek a more holistic change process through the middle-out approach. As we continue to study and implement our approaches, we hope to validate and refine them, providing productive starting points for change efforts at other institutions.

**Acknowledgements**

We thank the Association of American Universities and the Helmsley Charitable Trust for funding this work.